\documentclass[twocolumn,showpacs,preprintnumbers,amsmath,amssymb,superscriptaddress]{revtex4}

\usepackage{epsf}
\usepackage{graphicx}
\usepackage{sidecap}

\usepackage{color}

\def \beq {\begin{equation}}
\def \eeq {\end{equation}}
\begin{document}

\title{{Observation of Dirac state in half-Heusler material YPtBi}}

\author{M.~Mofazzel~Hosen}\affiliation {Department of Physics, University of Central Florida, Orlando, Florida 32816, USA}
\author{Gyanendra Dhakal} 
\affiliation {Department of Physics, University of Central Florida, Orlando, Florida 32816, USA}

\author{Klauss~Dimitri}\affiliation {Department of Physics, University of Central Florida, Orlando, Florida 32816, USA}


\author{Hongchul Choi}
\affiliation {Theoretical Division, Los Alamos National Laboratory, Los Alamos, New Mexico 87545, USA}










\author{Firoza Kabir}
\affiliation {Department of Physics, University of Central Florida, Orlando, Florida 32816, USA}

\author{Christopher Sims}
\affiliation {Department of Physics, University of Central Florida, Orlando, Florida 32816, USA}






	\author{Orest Pavlosiuk}
	\affiliation {Institute of Low Temperature and Structure Research, Polish Academy of Sciences,	50-950 Wroc{\l}aw, Poland}
	\author{Piotr Wi{\'s}niewski }
	\affiliation {Institute of Low Temperature and Structure Research, Polish Academy of Sciences, 50-950 Wroc{\l}aw, Poland}

\author{Tomasz~Durakiewicz}
\affiliation {Condensed Matter and Magnet Science Group, Los Alamos National Laboratory, Los Alamos, NM 87545, USA} 
\affiliation {Institute of Physics, Maria Curie - Sklodowska University, 20-031 Lublin, Poland}

\author{Jian-Xin~Zhu} 
\affiliation {Theoretical Division, Los Alamos National Laboratory, Los Alamos, New Mexico 87545, USA}
\affiliation{Center of Integrated Nanotechnologies, Los Alamos National Laboratory, Los Alamos, New Mexico 87545, USA}

\author{Dariusz Kaczorowski}
\affiliation {Institute of Low Temperature and Structure Research, Polish Academy of Sciences,
	50-950 Wroc{\l}aw, Poland}

\author{Madhab~Neupane}
\affiliation {Department of Physics, University of Central Florida, Orlando, Florida 32816, USA}

\date{18 June, 2013}
\begin{abstract}

{The prediction of non-trivial topological electronic states hosted by half-Heusler compounds makes them prime candidates for discovering new physics and devices as they harbor a variety of electronic ground states including superconductivity, magnetism, and heavy fermion behavior. Here we report normal state electronic properties of a superconducting half-Heusler compound YPtBi using angle-resolved photoemission spectroscopy (ARPES). Our data reveal the presence of a Dirac state at the $\Gamma$ point of the Brillouin zone at 500 meV below the chemical potential. We observe the presence of multiple Fermi surface pockets including two concentric hexagonal and six half oval shaped pockets at the $\Gamma$ and K points of the Brillouin zone, respectively. Furthermore, our measurements show Rashba-split bands and multiple surface states crossing the chemical potential which are supported by the first-principles calculations. Our finding of a Dirac state in YPtBi plays a significant role in establishing half-Heusler compounds as a new potential platform for novel topological phases and explore their connection with superconductivity.
	}

\end{abstract}
\date{\today}
\maketitle

\begin{center}
	I. Introduction
\end{center}

Topological quantum materials with non-trivial electronic band structures have gained intense research interest due to the possibility of exploiting exotic new physics such as Majorana fermions and the quantum spin Hall effect, as well as for their broad potential applications in quantum computing, spintronics, etc \cite{Hasan, SCZhang, Xia, Hasan_review_2}. These further lead to the realization and discovery of numerous topologically non-trivial states such as the topological insulators (TI), topological Kondo insulators, topological crystalline insulators, Dirac, Weyl, and nodal-line semimetals in various material families \cite{ Hasan_review_2, Bansil, Dai,  Neupane, Neupane_2, Neupane_1, TCI, Nagaosa, NdSb1, MH2, MH4,  Young_Kane,  Dai_LiFeAs,  Suyang_Science, Hong_Ding, TaAs_theory_1, TaAs_theory, MH3, MH1}.  Interestingly, ternary half-Heusler compounds have been theoretically predicted to provide a platform for realizing non-trivial electronic states as a result of their innate characteristic to obtain optimized parameters of topological order and topological phase transitions via tunable band gap and diverse spin-orbit coupling (SOC) \cite{ Ashvin, chadov, Lin, Bansil2, Ruan}. Furthermore, the half-Heusler compounds containing lanthanide elements with strongly correlated \textit{f}-electrons already have various ground states such as magnetism and superconductivity \cite{oster, f_elec}. Therefore, the presence of \textit{f}-electrons with these ground states in this system could lead to a new platform to study the correlated phenomena.\\~\\
  Despite numerous theoretical predictions, there are only a limited number of experimental studies which confirm non-trivial topology of band structure of half-Heusler compounds \cite{ Liu, Ong, Chen, Logan}. In 2016, large and negative longitudinal magnetoresistance has been reported as a signature of the chiral anomaly in GdPtBi \cite{Ong}. An angle-resolved photoemission spectroscopy (ARPES) study of RPtBi (R = Lu, Dy, Gd) by Liu \textit{et al.} \cite{Liu} has not found direct evidence of topological surface state in these materials. Recent ARPES studies of LnPtBi (Ln = Lu,Y) \cite{Chen} and LuPtSb \cite{Logan} have reported a topological surface state. However, the assignment of topological surface state as well as the origin of those states does not agree with each other. Therefore, it is desirable to study half-Heusler compounds in order to shed light on these issues.\\~\\
  \begin{figure*}
\centering
\includegraphics[width=18cm]{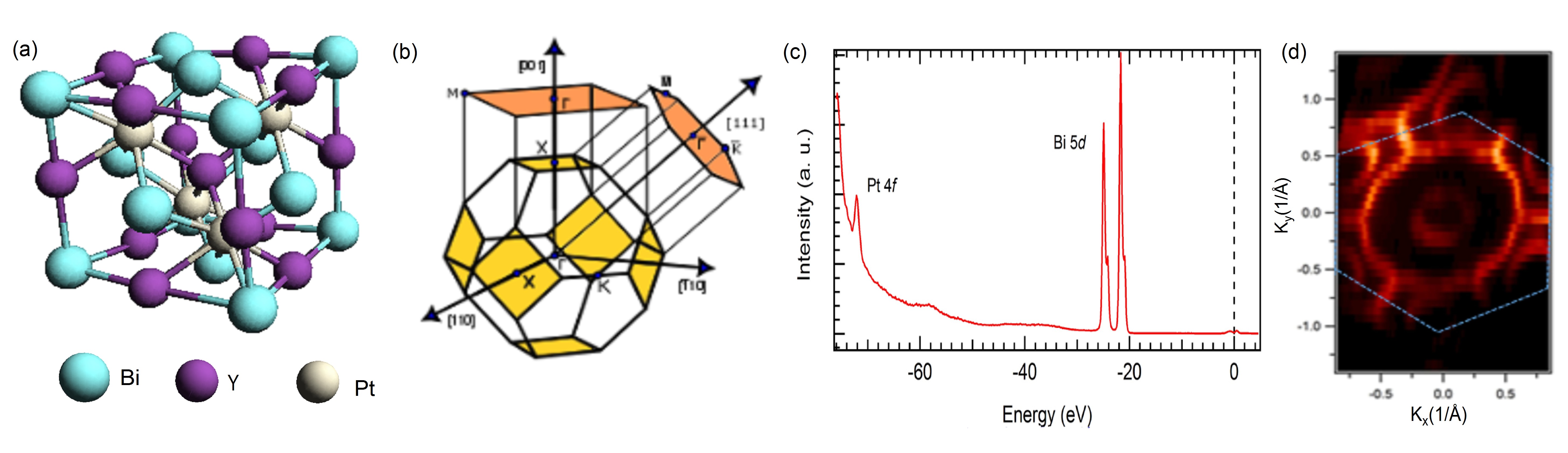}
\caption{{Crystal structure and sample characterization of YPtBi}.  (a) The crystal structure which is comprised of a zinc-blende
type crystal structure from the Pt and Bi atoms and a rocksalt type unit cell from Y and Bi.
(b) The bulk Brillouin zone with surface projections along the [001] and [111] directions. High symmetry points are marked in the plot. 
(c) Core level photo-emission spectrum. Sharp peaks of Pt 4\textit{f} and Bi 5\textit{d} are observed.
(d) Spectroscopically measured hexagonal Fermi surface of YPtBi. Dashed blue hexagon is to show the first Brillouin zone. Observation of hexagonal Brillouin zone confirms that the crystals cleaves along the (111) plane. Measurements were performed at ALS beamline 10.0.1 using a photon energy of 60 eV and at a temperature around 15 K. }
\end{figure*}

In this work, we report the detailed electronic structure study of YPtBi using ARPES and first-principles calculations. Our systematic electronic structure study reveals the hexagonal Fermi surface along with the presence of multiple Fermi pockets. Particularly, we find multiple Fermi surface pockets such as hexagonal and oval-shaped pockets around the zone center and the K points of the Brillouin zone (BZ), respectively. Interestingly, our data reveal a Dirac state at the $\Gamma$ point of the BZ where the Dirac point locates at 500 meV below the chemical potential which is further supported by our first-principles calculations. Moreover, we report Rashba-splitting in the vicinity of the Fermi surface at the M point of the BZ. We also report the presence of multiple surface states in this noble system. Our findings bolster the evidence for the existence of an exotic state within the half-Heusler family. As YPtBi is reported to be an intrinsic superconductor with \textit{T}$_c$ = 0.97 K \cite{SC, Butch, Nakajima, Tomasz}, our study serves a milestone to investigate the Dirac phase in superconductive YPtBi.\\


\begin{figure*}[ht]
\centering
\includegraphics[width=18.5cm]{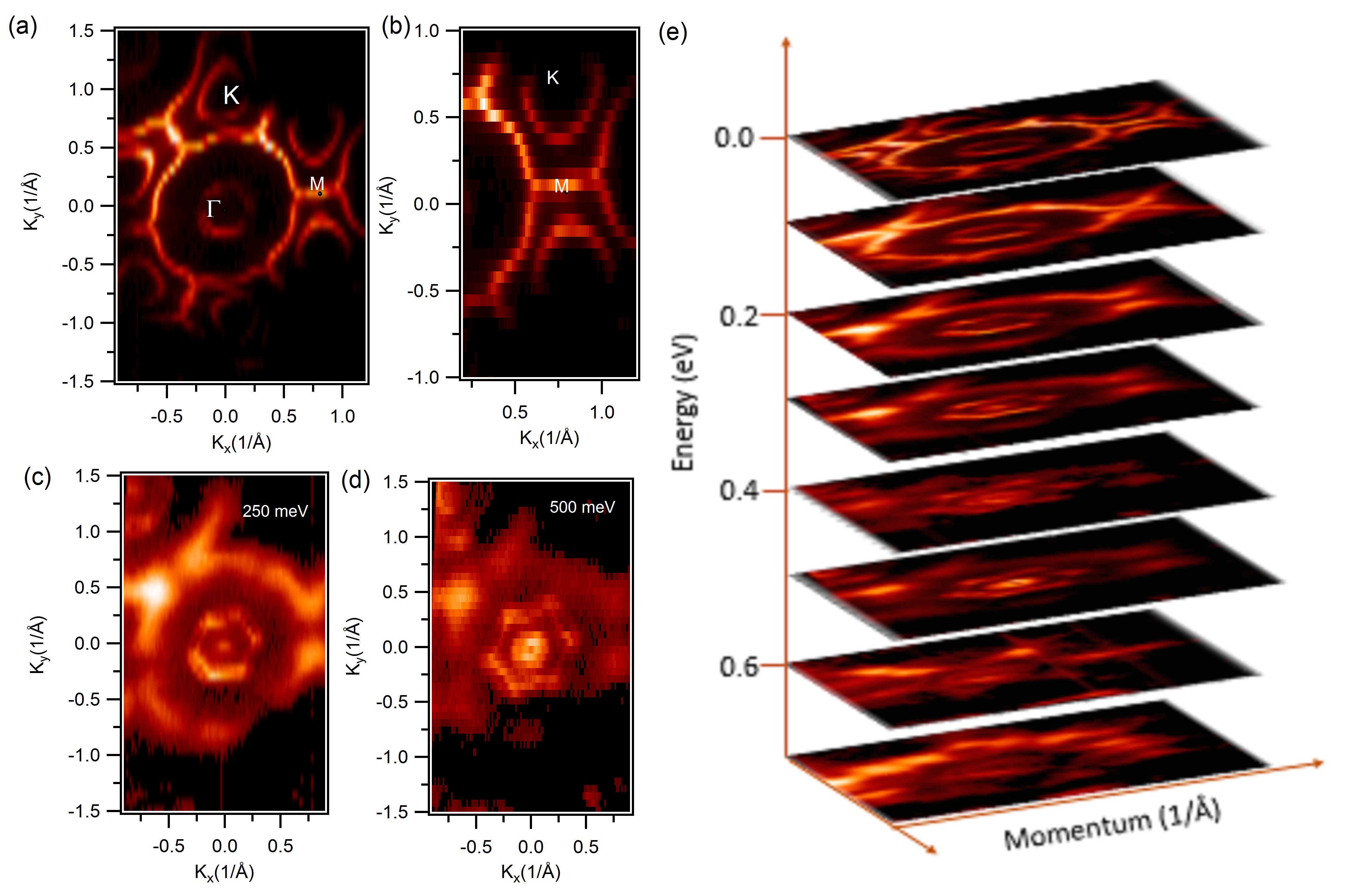}
\caption{{Fermi surface map of YPtBi.}  (a) Hexagonal Fermi surface map. High symmetry points are noted in the plot. (b) Zoomed view of Fermi surface map along the K-M-K direction. (c)-(d) Constant energy contour plots at different binding energies. Binding energies are noted in the plots. (e) Constant energy contour plots are stacked together in order to show the shape of the band. All the measurements were performed at ALS beamline 10.0.1 at a temperature of around 15 K. }
\end{figure*}

\begin{center}
	II. Method
\end{center}

The single crystals of YPtBi were grown from Bi flux as described in Ref. \cite{Tomasz}. The chemical composition of YPtBi was checked by energy-dispersive X-ray analysis using a FEI scanning electron microscope equipped with an EDAX Genesis XM4 spectrometer. Homogeneous single-phase with a stoichiometry close to equiatomic was observed.
Crystal structure of our sample was studied by means of X-ray diffraction on powdered single crystals using an X'pert Pro PAN analytical diffractometer with Cu-K$\alpha$ radiation. It was confirmed that YPtBi crystalizes in a space group $F\bar{4}3m$ with lattice parameter a = 6.65(1) $\AA$, which is in a perfect agreement with previously reported  data \cite{Butch}. The synchrotron based experiments were performed at the ALS BL 10.0.1 equipped with R4000 hemispherical electron analyzer at temperature of 15 K. For the synchrotron measurement the energy resolution was set to better than 20 meV and the angular resolution was set to better than 0.2$^{\circ}$. The electronic structure calculations were carried out using the full-potential linearized augmented plane wave (FP-LAPW) method implemented in the WIEN2k package \cite{39}, and the PBEsol was used as the exchange-correlation functional \cite{40}. In order to simulate surface effects on the (111) surface, we constructed the hexagonal unit cell of 3 formula units with c-axis normal to (111) surface. Then we built the 1 $\times$ 1 $\times$ 6 supercell for the (111) surface, with a vacuum thickness of 20 \AA. The spin-orbit coupling was considered as a perturbation in the electronic structure calculations.\\


\begin{center}
	III. Results and discussion
\end{center}
Generally, half-Heusler ternary compounds crystallize with space group $F \bar{4}3m$ ($\#$ 216) lacking inversion-center symmetry \cite{Cs}. Figure 1(a) shows the unit cell of YPtBi which consists of a zinc-blende type crystal structure from the Pt and Bi, and a rocksalt type unit cell from Bi and Y \cite{Butch, Tomasz}. The sample normally cleaves along either the (111) or (001) plane. Figure 1(b) shows the schematic bulk Brilloun zone with surface projections on the (001) and (111) plane. High symmetry points are also marked in the plot where $\Gamma$ is the center of the BZ, K is defined as the each corner of hexagon and M is the midpoint of two successive corner points. Figure 1(c) represents the spectroscopic core level measurement of YPtBi. Here, we observe the sharp core level peaks of Bi 5\textit{d} and Pt 4\textit{f} at around 26.5 eV and 71 eV, respectively. This confirms the high quality of the sample used for our experimental measurements. Furthermore, larger spectral weight of Bi 5\textit{d} indicates the Bi termination. An ARPES measured hexagonal-shaped Fermi surface of YPtBi is seen in Fig. 1(d) which further confirms that the sample was cleaved along the (111) plane. The blue hexagon is a guide for the eye representing the first BZ..\\~\\ 

In order to unveil the detailed electronic structure, we present our experimental data in Figs. 2 and 3. Figure 2(a) shows the Fermi surface map revealing the presence of multiple Fermi pockets including two circular-like Fermi pockets around the $\Gamma$ point and six half oval-shaped pockets around the K points of the BZ. The high symmetry points are marked in the plot. Figure 2(b) presents the zoomed in plot along the K-M-K high symmetry direction that clearly shows the half oval-like Fermi pockets. Figure 2(c) shows the constant energy contour plot at the binding energy of 250 meV where the circular-like Fermi pocket at the Fermi surface around zone center point evolves into a twin hexagonal shape. This feature is a consequence of the possible spin-splitting, commonly seen on the surface of the non-centrosymmetric metallic compounds. Futhermore, a small circular feature evolves at the zone center point as a result of metallic band crossings. Figure 2(d) shows a constant energy contour in the vicinity of the Dirac point ( binding energy $\sim$ 500 meV) in which we observe the circular-shaped feature splitting into two concentric circles. In order to reveal the nature of the Fermi pockets, constant energy contour plots at various binding energy are taken and stacked into an energy vs momentum plot in Fig. 2(e). By moving towards the higher binding energy, we observe that the circular-like pocket at the zone center increases in size and resembles a perfect hexagonal shape. This confirms both the hole-like nature and hexagonal shape of the pocket. Additionally, the oval-like shape around the K point shrinks into a point and a small circular feature evolves at the zone center around 250 meV. This confirms the electron like nature of the Fermi pockets at the K point of the BZ. Moreover, at 500 meV below the chemical potential, the complex oval-like feature takes shape which indicates the approximate position of the Dirac point. At the Fermi level we observe a line-like feature along the M points connecting the two consecutive BZ as well as separating the two oval-like pockets. However, as we move towards higher binding energies it is observed that the line gap increases and evolves into two distinct segments breaking the outer circle around the $\Gamma$-point. The special outer circle located at around 100 meV below the Fermi surface could potentially represent another pair of Kramers points \cite{Liu}. \\~\\

\begin{figure*}
	\centering
	\includegraphics[width=18cm]{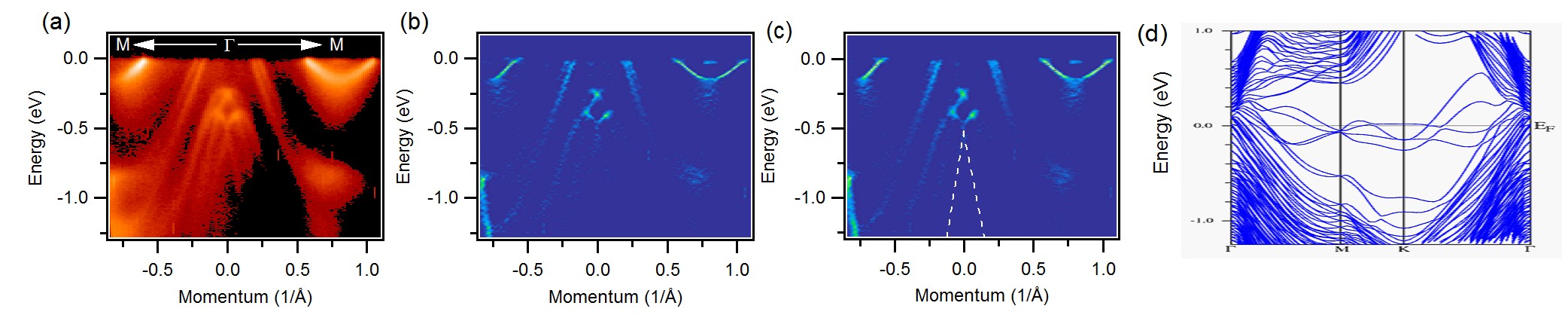}
	\caption{{Observation of Dirac state in YPtBi.}
		(a) Measured dispersion map along the M-$\Gamma$-M direction at a photon energy of 60 eV. (b)-(c) Second derivative plot of Fig.(a) using the curvature method \cite{2D}. White dashed line in Fig. (c) serves as a guide for the eyes. (d) Theoretical energy dispersion map along the $\Gamma$-M-K-$\Gamma$ high symmetry direction. All the experiments were performed at the ALS end-station 10.0.1 at a temperature of around 15 K.}
\end{figure*}

In order to reveal the energy location of the Dirac point, we present our band dispersion map in Fig. 3. Figure 3(a) shows the dispersion map along the M-$\Gamma$-M direction in which we observe hole-like bands at the $\Gamma$ point and electron-like bands at the M point crossing the Fermi level. Interestingly, we observe a Dirac state at the $\Gamma$-point of the BZ at 500 meV below the chemical potential. Furthermore, we observe another surface state at around 250 meV below the Fermi level. Figure 3(b),(c) presents the second derivative plot using curvature method where the Dirac point is clearly visible. In order to reveal the origin of the bands, slab calculations were performed which are presented in Fig. 3(d). Here, it is clearly observed that the electron-like bands cross the chemical potential. Furthermore, we observe the Rashba-split states in the vicinity of the chemical potential around the M point. The surface calculations (see Fig. 4(a)) confirm the surface nature of Rashba-split states and the electron-like band at the M-point.\\~\\

 \begin{figure}
 	\centering
 	\includegraphics[width = 8.7cm]{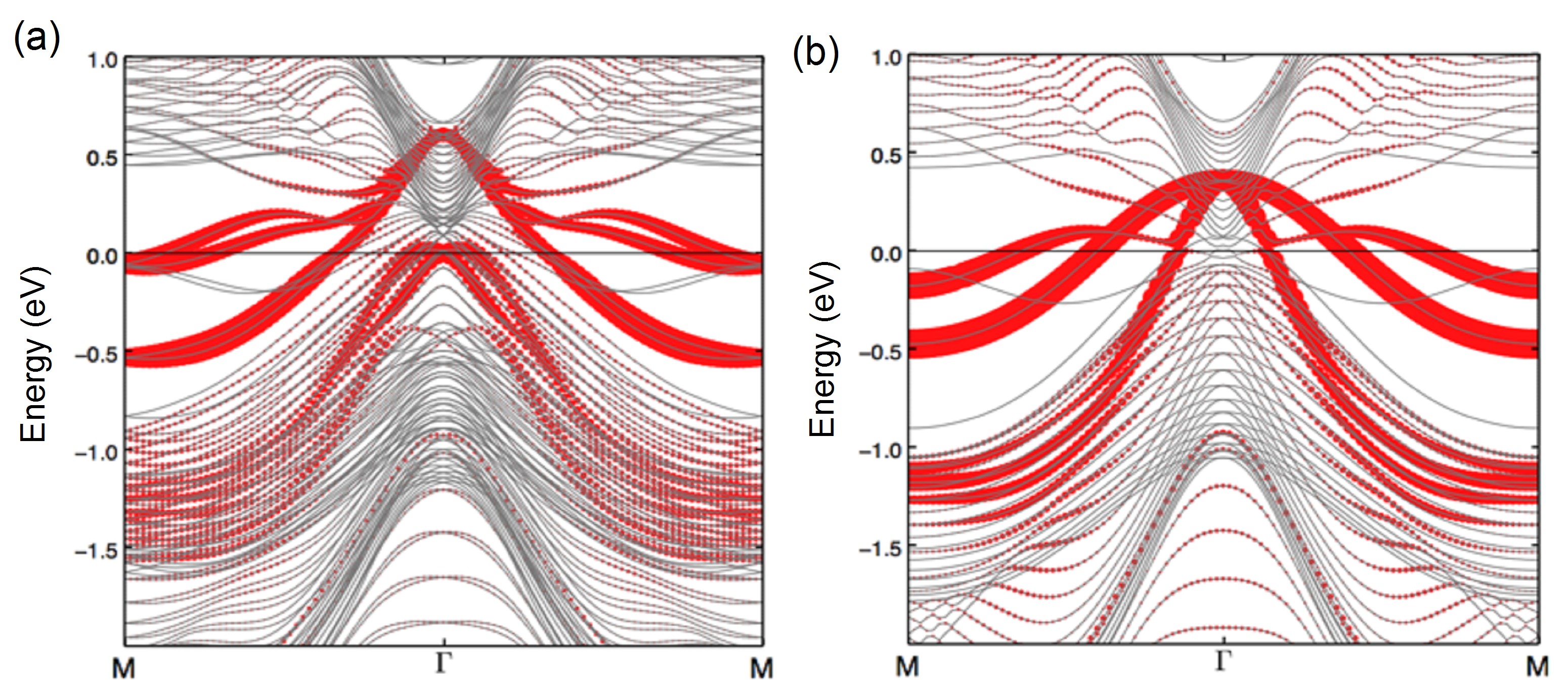}
 	\caption{{First-principles calculations of Bi-terminated surface state of YPtBi.} 
 		(a) First-principles calculations with inclusion of spin-orbit coupling (SOC) along the M-$\Gamma$-M direction. 
 		(b) First-principles calculations without inclusion of SOC along the M-$\Gamma$-M direction. The red highlight indicates the surface states.
 	}
 \end{figure}
 
Here, we discuss our first-principles calculations as depicted in Fig. 4. Figure 4(a) shows the calculations of electronic structure for the Bi-terminated surface with the inclusion of spin-orbit coupling along the M-$\Gamma$-M direction. Here, red highlighted bands represent the surface bands. The calculations agree well with our measurements which establish the Dirac point located at around 500 meV below the chemical potential. Furthermore, it shows the Rasba-split band and the electron-like band around the M point. The energy location of the other surface state at the $\Gamma$ point is slightly mismatched with our experimentally observed energy location. Since these bands are the metallic surface states defined within the 2D surface, their positions might be controlled by impurity potential at the surface. This type of energy mismatching has also been reported earlier \cite{Chen}. Moreover, electron-like band along the $\Gamma$-M direction is confirmed to be surface originated. Figure 4(b) displays the first-principles calculations of electronic structure for the Bi-terminated surface without the inclusion of SOC, which does not show a Dirac-like state. We conclude that the linear Dirac bands originate from the spin-orbit coupling arising due to heavy Bi atom.\\~\\

Considering the experimental and theoretical study of half-Heulser compound YPtBi, both confer the same information regarding the location of the Dirac points 500 meV below the chemical potential at the $\Gamma$ point. At this high symmetry point, we notice that the Dirac state originates from the surface which is the result of the strong SOC. 
We note that our data does not show any intrinsic global band gap in YPtBi. Therefore, we attribute such linearly dispersive band to a Dirac state instead of topological surface state. Moreover, our experimental measurements as well as calculations reveal possible Rashba-split states in the vicinity of Fermi surface at the M point due to an antisymmetric spin-orbit coupling created by the electric field gradient as a consequence of non-centrosymmetric geometry. \\~\\
\begin{center}

	IV. Conclusions
\end{center}
In conclusion, we have performed ARPES measurements on the ternary superconducting half-Heusler compound YPtBi in its normal state. Our data reveal the presence of multiple Fermi pockets at the Fermi surface of YPtBi. Furthermore, we directly observe the Dirac state at 500 eV below the chemical potential at the $\Gamma$ point of the BZ. We further observe multiple surface states in our Fermi surface maps. Our first-principles calculations reveal a Rashba-split feature at the M point. Our study will stimulate the research interest to investigate the Dirac phase in a intrinsic superconducting YPtBi and exotic states in other half-Heuslers.



\bigskip
\begin{center}
	Acknowledgments
\end{center}

M.N. is supported by the start-up fund from University of Central Florida.
T.D. is supported by NSF IR/D program. 
O.P., P.W. and D.K. is supported by the National Science Centre (Poland) under research grant 2015/18/A/ST3/00057. Work at Los Alamos National Laboratory (LANL) is supported by U.S. DOE Contract No. DE-AC52-06NA25396 through the LANL LDRD Program (H.C. and J.-X.Z.). This work was, in part, supported by the Center for Integrated Nanotechnologies, a U.S. DOE BES user facility, in partnership with the LANL Institutional Computing Program for computational resources.
 We also thank Sung-Kwan Mo for beamline assistance at the LBNL.

\*Correspondence and requests for materials should be addressed to M.N. (Email: Madhab.Neupane@ucf.edu).

\end{document}